# On age of 6070 Rheinland and 54827 (2001 NQ8) asteroid pair

Rosaev A., Plavalova E

Research and Educational Center "Nonlinear Dynamics", Yaroslavl State University
E-mail: hegem@mail.ru
Astronomical Institute Slovak Academy of Sciences

In this paper we present results of our studying of famous very young pair of asteroids 6070 Rheinland and 54827 (2001 NQ8). We have done numeric integration of orbits of pair with only planet perturbations and include Ceres and Vesta effect. We have confirmed results of previous studying, obtained with different integrators. And we confirm significant effect of Ceres and Vesta perturbation on dynamic of this pair. We find that effect of other massive asteroids is insignificant. According our results, more probable age of 6070 Rheinland and 54827 (2001 NQ8) pair is 16.2 kyrs. Our value of age is very close to most recent age determination by Vokrouhlicky et al [12], obtained with different method. After the compare our results, we can conclude, that non-gravitational forces are small and large number of clones is not necessary in studying of this pair.

As an additional way of studying of close orbits dynamics, we calculate relative velocity in pair during numeric integration. Normal component of velocity show a very good convergence at epoch of closest encounter in pair.

Finally, we note the very small inclination of orbits of both members of pair just after forming event 15 kyrs ago. This fact is necessary to account at precision breakup reconstruction. We present a simple analytic model, explained observed orbital elements evolution.

Introduction

The discovery and very detail studying of pair 6070 Rheinland and 54827 (2001 NQ8) was published by Vokrouhlicky & Nesvorny [8, 10, 11]. They calculate age of forming breakup in this pair $17.22\pm0.28$ kyr ago. However, they not take into account Ceres and Vesta perturbations. Late Galad [4] develop integration with perturbation of 256 most massive asteroids and obtain for age of this pair value 16.2 kyrs. Our goal is to study, how significant this difference and which asteroids takes main contribution on it.

Vokrouhlicky & Nesvorny [11] also find extremely small median encounter velocity at infinity: 17 cm s$^{-1}$, while its component normal to the orbit of larger asteroid (6070) Rheinland is only 21 mm s$^{-1}$.

The proper orbital elements of pair are very close (table 1). As noted by Vokrouhlicky & Nesvorny, these two orbits have identical proper elements within the statistical uncertainty. However, for so young pair we can use osculating elements (table 2).

Throughout the article we have used standard notations for orbital elements $a$ – semimajor axis in a.u., $e$ – eccentricity, $i$ – inclination, $\Omega$ – longitude of ascending node, $\omega$ – perihelion argument, $\varpi$ – longitude of perihelion (the angular elements are in degrees).

Table 1. Synthetic proper orbital elements of pair 6070 Rheinland and 54827 (2001 NQ8) [5]

| asteroid | g, arcsec/yr | s, arcsec/yr | Sin($i$) | $e$ | $a$ |
|---|---|---|---|---|---|
| 6070 Rheinland | 39.9733 | -44.8937 | 0.0379871 | 0.176155 | 2.38785 |
| (54827) 2001 NQ8 | 39.9712 | -44.8917 | 0.0379912 | 0.176144 | 2.3878 |

Table 2 Osculating orbital elements of the pair 1998-Jul-06 (JD2451000.5)

| | $\omega$ | $\Omega$ | $i$ | $e$ | $a$ |
|---|---|---|---|---|---|
| 6070 Rheinland | 292.43979 | 84.128894 | 3.1328 | 0.210202 | 2.3882 |
| 54827 (2001 NQ8) | 292.40110 | 84.106397 | 3.1324 | 0.210198 | 2.3876 |

**Method**

We repeat numeric integration of this pair using integrator, which is differs than in paper [11] (Vokrouhlicky & Nesvorny 2009). To study the dynamical evolution of these close asteroid pairs, the equations of the motion of the systems were numerically integrated 50 kyrs into the past, using the N-body integrator Mercury (Chambers, 1999)[1] and the Everhart integration method (Everhart, 1985)[3]. On base of [4,11] age estimation, we expect that this time interval is sufficient. We made three series of integration. In the first we use only large planets perturbations. In the second we add Ceres, Vesta, Juno and Pallas. Finally, we add perturbations of some other large asteroids: Hygeia, Interamnia, Davida, Eunomia (table 3).

We have not taken into account any non-gravitational forces, because they poorly known and have no time to sufficient change of orbital elements in considered pairs.

**Table 3.** Osculating orbital elements and masses of large asteroids. Epoch 16.01.2009 (JD 2454847.5)

|           | $m/m_{sun}$ [2,6] $10^{-10}$ | $\omega$   | $\Omega$   | $i$      | $e$      | $a$      |
|-----------|------------------------------|------------|------------|----------|----------|----------|
| Ceres     | 4.73                         | 73.799349  | 80.501619  | 10.58294 | 0.077887 | 2.766180 |
| Pallas    | 1.07                         | 309.852968 | 173.212145 | 34.85118 | 0.230889 | 2.776424 |
| Vesta     | 1.33                         | 149.799988 | 103.961308 | 7.134869 | 0.090298 | 2.360763 |
| Juno      | 0.15                         | 247.975605 | 170.181889 | 12.96766 | 0.25787  | 2.669314 |
| Hygiea    | 0.47                         | 314.616094 | 283.688404 | 3.843917 | 0.119662 | 3.136872 |
| Interamnia| 0.35                         | 94.312521  | 280.689537 | 17.32091 | 0.146025 | 3.064568 |
| Davida    | 0.3                          | 338.899084 | 107.788255 | 15.93802 | 0.182459 | 3.171485 |
| Eunomia   | 0.16                         | 96.868528  | 293.534723 | 11.74861 | 0.186715 | 2.644412 |

To study convergence in this pair we use direct determination of distance at close encounter close to 17 kyrs ago and components of relative velocity calculation along our backward numeric integration.

We use Gauss equations for orbital elements variations [7]:

$$\delta a = \frac{2}{n\eta}\left[V_R e \sin f + V_T(1 + e\cos f)\right]$$

$$\delta e = \frac{\eta}{na}\left[V_R \sin f + V_T \frac{e + 2\cos f + e\cos^2 f}{1 + e\cos f}\right]$$

$$\delta i = \frac{\eta}{na} V_Z \frac{\cos(\omega + f)}{1 + e\cos f} \qquad (1)$$

$$\delta\Omega = \frac{\eta}{na \sin i} V_Z \frac{\sin(\omega + f)}{1 + e\cos f}$$

$$\delta\varpi = \frac{\eta}{nae}\left(-V_R \cos f + V_T \sin f \frac{2 + e\cos f}{1 + e\cos f}\right) + 2\sin^2\frac{i}{2}\delta\Omega$$

$$\eta = (1 - e^2)^{-1/2}$$

These equations may be inverted to obtain values of relative velocity at each step of numeric integration. We use z-component of relative velocity as a low dependent on short periodic perturbations:

$$V_Z^2 = (1 - e^2)(1 + e\cos(f))^2 (GM/a)\left[(\delta i)^2 + (\delta\Omega)^2 \sin^2 i\right] \qquad (2)$$

For the control and for other components of velocity estimation we use simplified formula (Nesvorny & Vokrouhlicky, 2006)[8]:

$$\Delta V(t) = na\sqrt{k_1(\sin i \Delta\Omega)^2 + k_2(e\Delta\varpi)^2} \qquad (3)$$

**Results**

The main results of our integration are in table 4 and fig. 1-2. In the first our computations, when only planetary perturbations have used, we obtain value of age of pair very close to Vokrouhlicky & Nesvorny [8,11] (17.33 against 17.22 kyrs). Note, that we do it without any clones and do not take into account Yarkovsky effect. There are some followings: 1) non-gravitational effects are insufficient in dynamic of considered pair, 2) integrator Mercury can be used for studying young asteroid families dynamic.

After that, we have include Ceres and Vesta perturbations and obtain for age of pair 16.2 kyrs, which is in agreement with Galad result [4] and notable smaller than value 17 kyrs for age in [11]. This result is in good agreement with result [4] and most recent result [12] (table 4).

Our third integration show, that other large asteroids have insignificant effect on considered pair dynamics.

The calculations of z-component of relative velocity along integrations prove these results (fig.2). In general, we have confirmed one of main results of (Vokrouhlicky & Nesvorny 2009) - very small relative velocity (about 20 cm/s) at encounter. Moreover, we can note, that vector of breakup velocity is close to tangential direction (along orbit).

However, it is evident, that perturbation of pair by large asteroids is important for this case. The effect of Ceres and Vesta is different on members of pair. The variations in true anomaly, which is affected by Ceres and Vesta perturbations at epoch of breakup near 17 kyrs ago are $0.4^o$ for 54827 and $3.5^o$ for Rheinland. This value is small, but not negligible at precision breakup process reconstruction.

In this paper we not separate Ceres and Vesta perturbations. But in according to Tsirvoulis and Novakovic result, [9] Vesta is a more significant perturber for pair 6070 Rheinland / 54827 (2001 NQ8) and secular resonance with Vesta may be important.

Table 4. Age estimations of 6070 Rheinland 54827 (2001 NQ8) pair

| Dmin, au | Age, kyrs | | Perturbation |
|---|---|---|---|
| Present paper | Present paper | Referenced | |
| 0.00007038 | -17.33 | -17.22[11] | Only planets |
| 0.00008162 | -16.19 | -16.33 [12] | CVJP |
| 0.00008515 | -16.21 | -16.2 [4] | CVJPHIDE |

Table 5. Relative velocities at epoch of pair origin

| Epoch, kyr | Vz,m/s | Vt,m/s | Vr,m/s | |
|---|---|---|---|---|
| -16.1920 | 0.01270597 | 0.20721435 | 0.03899169 | |
| -16.2040 | 0.01256533 | 0.18905496 | 0.00883932 | CVJP |
| -16.2045 | 0.01250731 | 0.19780865 | 0.02093736 | |
| -16.2060 | 0.01241232 | 0.19777214 | 0.02099078 | |

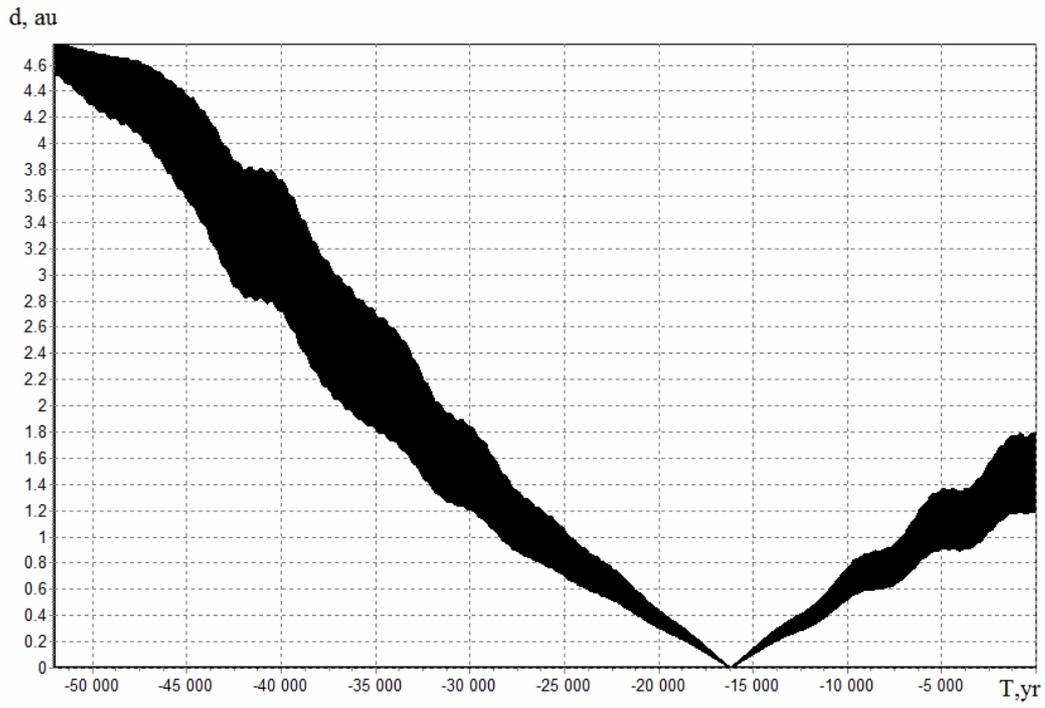
Fig.1. Distance between 6070 Rheinland and 54827 (2001 NQ8) evolution

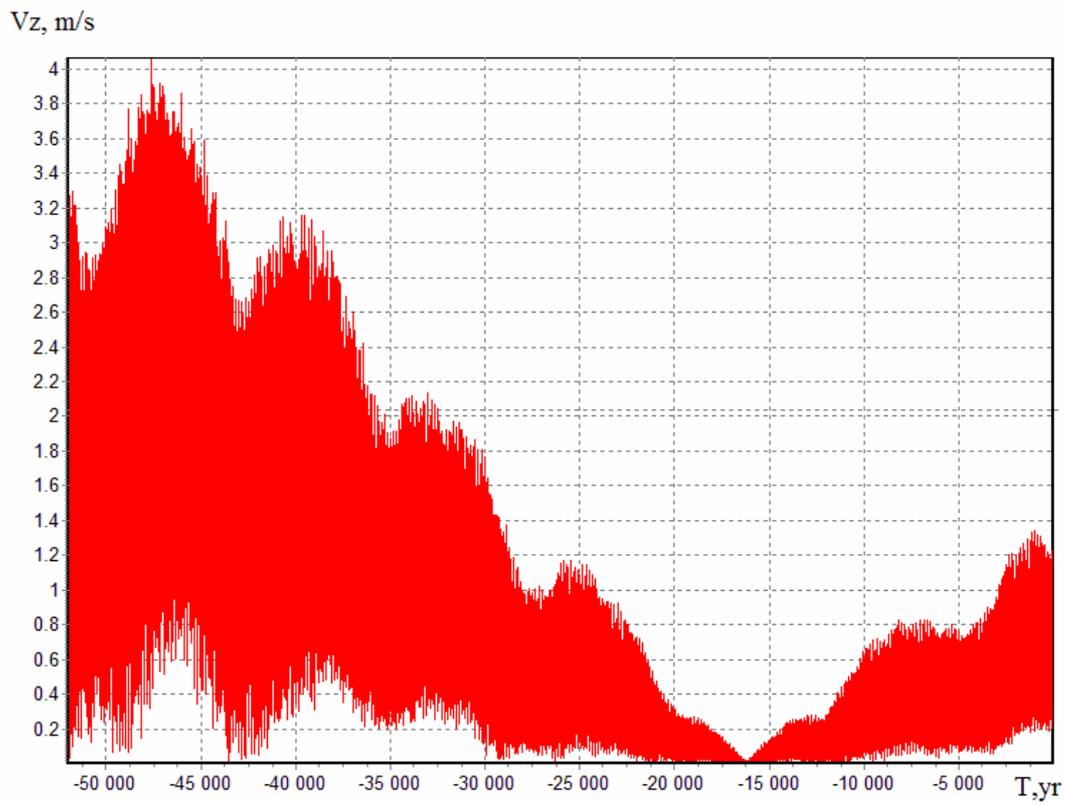
Fig.2 z-component of relative velocity evolution

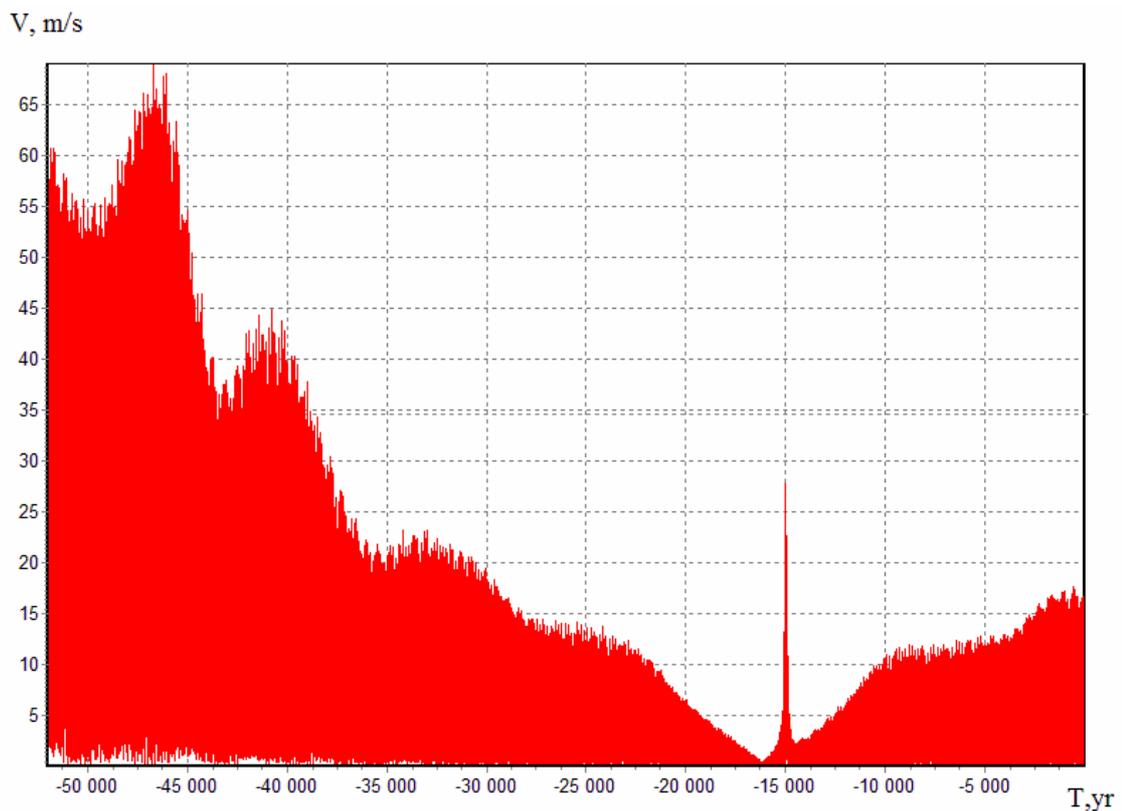

Fig.3 Averaged velocity, calculated by expression (3)

Conclusions

In this paper we present results of our studying of famous very young pair of asteroids 6070 Rheinland and 54827 (2001 NQ8).  We develop numeric integration of orbits of pair with only planet perturbations and include Ceres and Vesta effect. We have confirmed results of previous studying, obtained with different integrators. And we confirm significant effect of Ceres and Vesta perturbation on dynamic of this pair. We have found, that effect of other asteroids is insignificant. According our results, more probable age of 6070 Rheinland and 54827 (2001 NQ8) pair is 16.2 kyrs.

As a new method of studying of close orbits dynamics, we calculate relative velocity in pair during numeric integration. Normal component of velocity show a very good convergence at epoch of closest encounter in pair.